\documentclass{iau} 
\usepackage{graphicx}
\usepackage{subfig}

\title[Symposium 320.~~Stellar Wind -- Magnetosphere Interactions in Hot Jupiters] 
{Stellar Wind -- Magnetosphere Interactions in Hot Jupiters}

\author[Derek L. Buzasi]   
{Derek L. Buzasi$^1$}

\affiliation{$^1$Dept. of Chemistry and Physics, Florida Gulf Coast University \\
10501 FGCU Boulevard S., Fort Myers, FL 33965 USA \\ email: {\tt dbuzasi@fgcu.edu}}

\pubyear{2015}
\volume{320}  
\jname{Solar and Stellar Flares and Their Effects on Planets}
\editors{A.G. Kosovichev, S.L. Hawley, \& P. Heinzel, eds.}
\begin{document}

\maketitle

\begin{abstract}
One potential star-planet interaction mechanism for hot Jupiters involves planetary heating via currents set up by interactions between the stellar wind and planetary magnetosphere. Early modeling results indicate that such currents, which are analogous to the terrestrial global electric circuit (GEC), have the potential to provide sufficient heating to account for the additional radius inflation seen in some hot Jupiters. Here we present a more detailed model of this phenomenon, exploring the scale of the effect, the circumstances under which it is likely to be significant, implications for the planetary magnetospheric structure, and observational signatures.
\end{abstract}

\firstsection 
\section{Introduction}

Hot Jupiters frequently have radii which are “inflated” relative to the expected mass-radius relation 
\cite[(Fortney \& Nettleman 2010)]{Fortney10}. The degree of inflation is correlated with orbital semimajor axis (Demory \& Seager 2011) and typically becomes significant for $F > 2 \times 10^8 \rm ~erg~cm^{-2}~s^{-1}$, which corresponds to approximately the Alfv\'{e}n radius $a = r_{A}$, and there is some indication that it exhibits a dependence on stellar activity level \cite[(Buzasi 2013)]{Buzasi13}. Proposed models for this behavior include tidal heating \cite[(Gu et al. 2004)]{Gu04} and “Ohmic heating” due to interactions between atmospheric flows and the planetary magnetic field \cite[(Batygin et al. 2011)]{Batygin11}. \cite[Buzasi (2013)]{Buzasi13} proposed interactions between the stellar wind and the planetary magnetic field as an alternate mechanism; in this work we explore the ramifications of that suggestion in the context of a more complete planetary model.

\vspace*{-0.25 cm}
\section{Model and Results}

The model calculates currents internal to the planet generated by the electric potential difference produced by the stellar wind/magnetosphere interaction and mapped down to the outer layers of the planet, analogous to the
“global electric circuit” (GEC; \cite[Tinsley et al. 2007]{Tinsley07}). A solar wind model \cite[(Suzuki 2006)]{Suzuki06} is taken as input, and the conductivity is calculated for the planetary interior, which in turn is calculated using the MESA code \cite[(Paxton et al. 2011)]{Paxton11}. The presence of a $\sim 10 \rm G$ planetary magnetic field renders conductivity inhomogeneous. The magnetic dipole moment is presumed coaligned with the rotation axis and ecliptic poles, and the potential difference drives Pedersen and parallel (field-aligned) currents, resulting in Joule heating of the planetary interior.

A series of models with solar composition were calculated for $M_{PL} = 10^{-3} M_{\odot}$ ($\sim M_J$) using MESA \cite[(Paxton et al. 2011)]{Paxton11} and evolved to $t = 5 \rm~Gyr$. Model electron densities were combined with the adopted $10 \rm~G$ dipole planetary magnetic field to calculate the classical ($\sigma_0$), Pedersen ($σ\sigma_P$), and Hall ($\sigma_H$) conductivities. The ionospheric electric potential was taken as zero except at two regions in each hemisphere located at invariant latitude $\Lambda = \cos^{-1} \sqrt{\frac{R_{PL}}{R_M}}$ and separated by $\pi / 2$ in longitude. Regions were circular with diameter $\Lambda$. The potential adopted was the minimum of the wind-induced field $E = v_w B_w R_M$, or the value leading to the maximum energy deposition possible \cite[(Akasofu 1981)]{Akasofu81}, $\epsilon = v_w B_w^2 R_M^2$.

Heating resulting from these potential/conductivity combinations was calculated and incorporated into the
MESA models. The resulting energy deposition profiles for two typical models are shown in Figure 1a. Total heating in both cases is limited by the power available from the wind rather than by the wind-induced field strength. Figure 1b illustrates planetary interior models resulting from the two cases; note the growth of the 1-bar planetary radius from $\approx 1.05 R_J$ at 1 AU to $\approx 1.58 R_J$ at 0.014 AU. The latter corresponds to a stellar irradiance of $F = 7 \times 10^9 \rm~erg~cm^{-2}~s^{-1}$ and an orbital period of 0.6 days.

\begin{figure}[!tbp]
  \centering
  \subfloat[]{\includegraphics[width=0.4\textwidth]{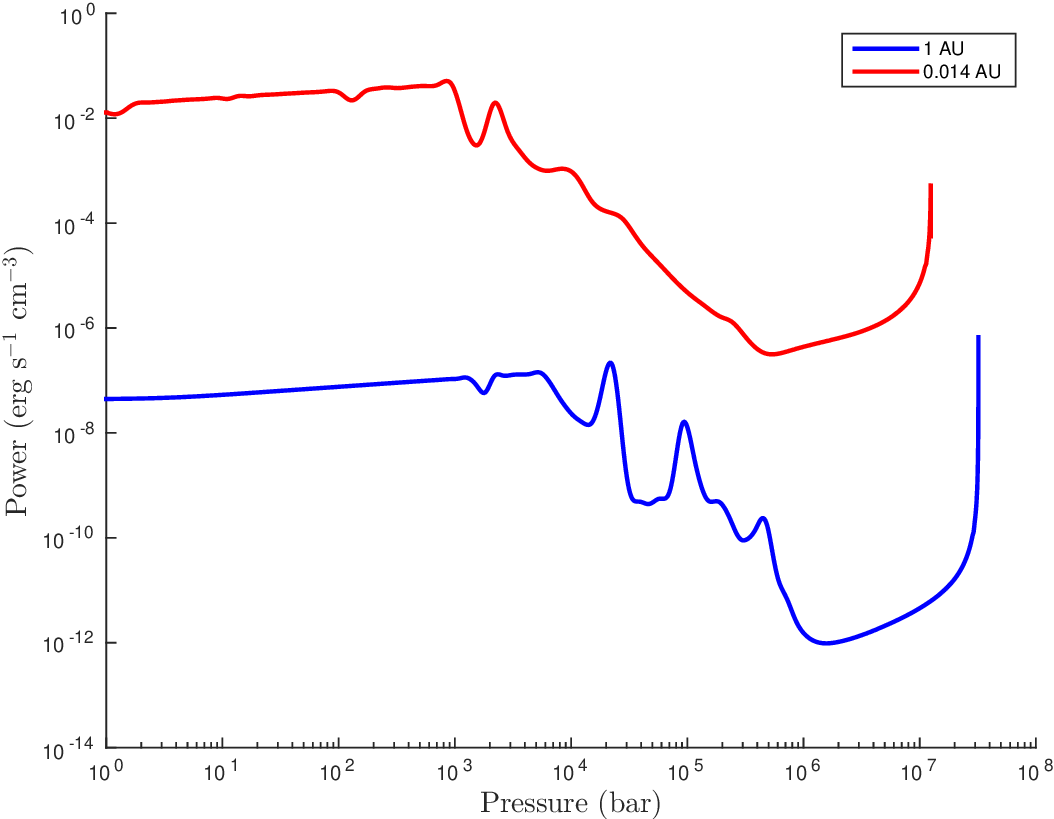}\label{fig:f1}}
  \hfill
  \subfloat[]{\includegraphics[width=0.4\textwidth]{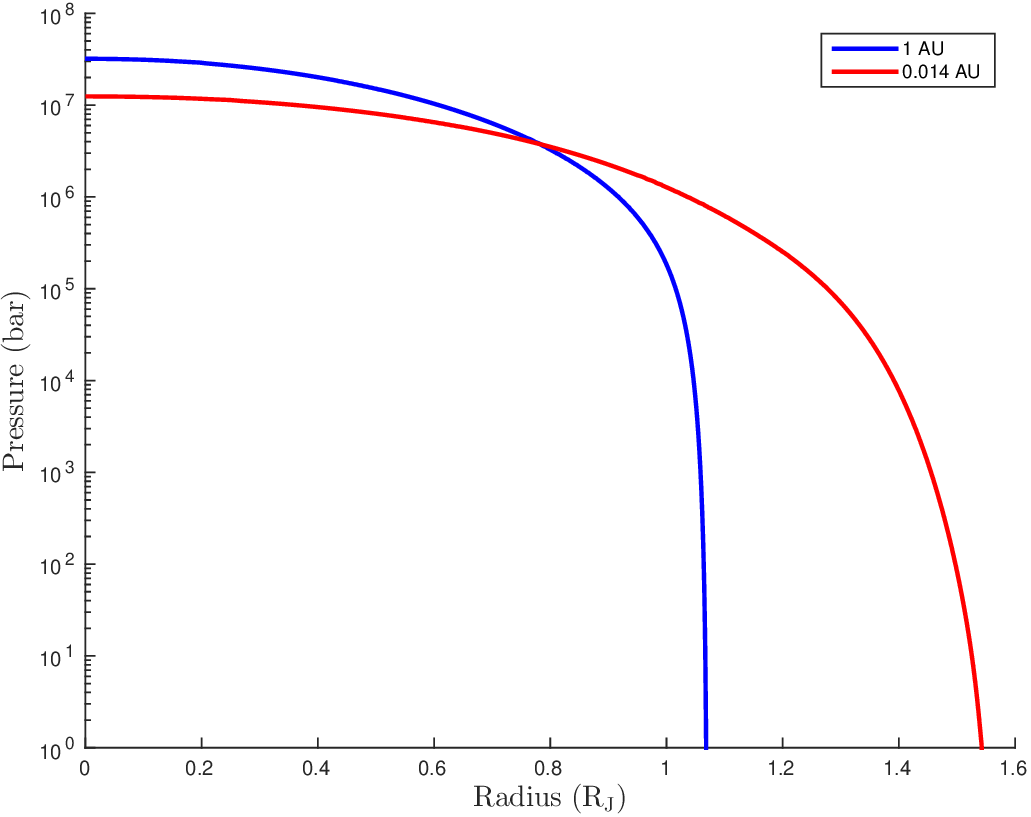}\label{fig:f2}}
  \caption{The left panel shows internal electric heating input due to GEC as a function of internal pressure for models with semimajor axes $a = 1 \rm~AU$ (blue) and $a = 0.014 \rm~AU$ (red). Maximum heating occurs at $r/R_{PL} = 0.973$ at $1 \rm~AU$ and at $r /R_{PL} = 0.945$ at $0.014 \rm~AU$. The right panel illustrates MESA models incorporating GEC heating for the same models, both with solar composition and $B = 10 \rm~G$, and incorporating stellar irradiation via a gray atmosphere.}
\end{figure}

\vspace*{-0.25 cm}
\section{Discussion}

Successful models which account for the observed radius excess in hot Jupiters by additional heating must be
capable of supplying $>10^{27} \rm~erg~s^{-1}$ to the convective portion of the planetary interior. The proposed GEC model is capable of such heating, and produces planetary radii broadly in accord with observations (Figure 2). Variations in planetary mass, planetary composition (leading to changes in conductivity), planetary magnetic field, and stellar magnetic field may account for the range of radii observed in the sample. In addition, we note that it is likely that multiple heating models coexist, including tidal heating (for noncircular orbits) and Ohmic heating (\cite[Batygin \& Stevenson 2010]{Batygin10}, \cite[Batygin et al. 2011]{Batygin11}).

Observational testing of the model is possible. In particular, predictions include
\begin{enumerate}
\item{The correlation of radius inflation with stellar magnetic field, potentially derivable from accurate stellar
activity proxies such as Ca II h+k and/or starspot coverage.}

\item{Planetary temperatures in excess of those possible based solely on radiative equilibrium calculations. Precise photometric transit observations and comparisons of atmospheric spectra to models should enable such testing.}
\end{enumerate}

\begin{figure}[!tbp]
  \centering
  \subfloat[]{\includegraphics[width=0.4\textwidth]{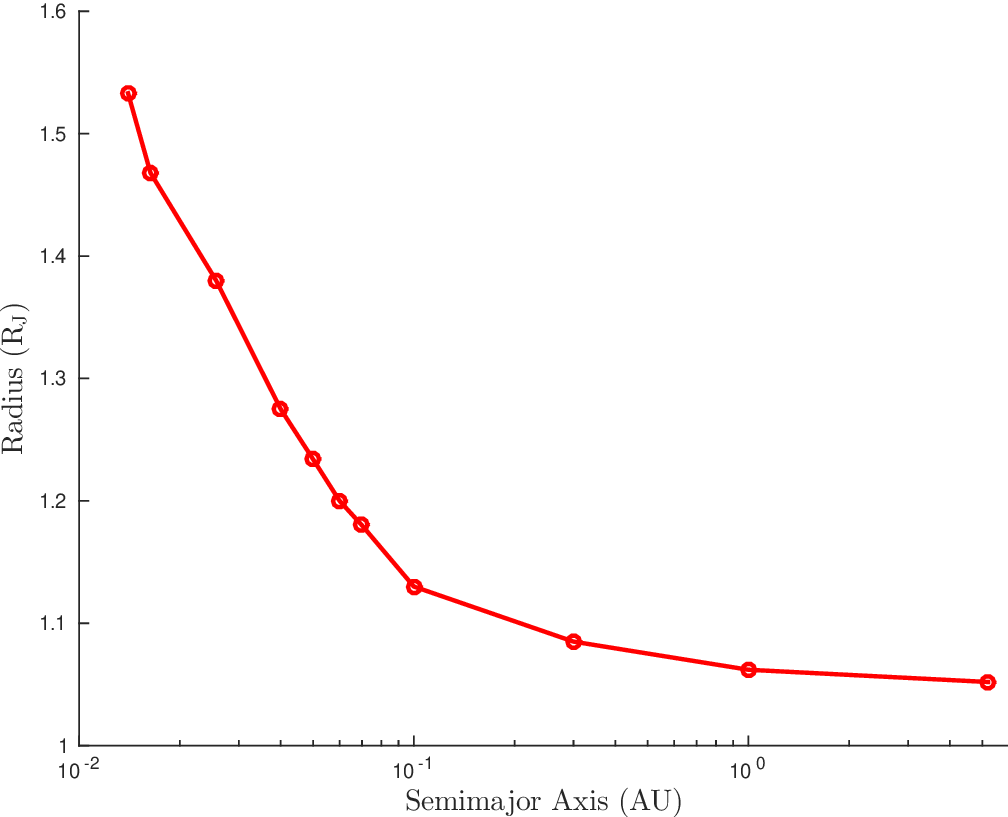}\label{fig:f3}}
  \hfill
  \subfloat[]{\includegraphics[width=0.4\textwidth]{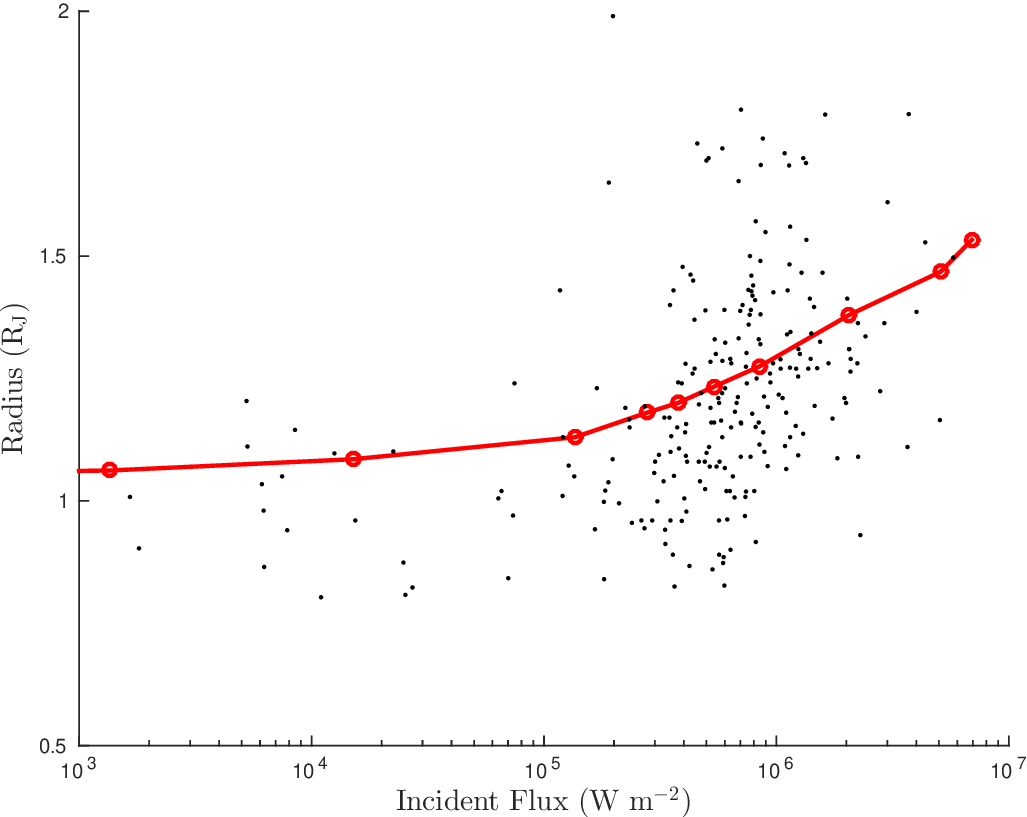}\label{fig:f4}}
  \caption{Left panel: planetary radii at $P = 1 \rm~bar$ for models at varying distances from the Sun. Heating leads to significant radius inflation for models with semimajor axes less than $\sim 0.1 \rm~AU$. All models were evolved to $t = 5 \rm ~Gyr$. Right panel: model radii plotted against stellar irradiation and compared to all known hot Jupiters ($R_{PL} > 0.8R_J$). All models assume $10\rm~G$ planetary magnetic fields.}
\end{figure}

Figure 2a summarizes model results over a range of semimajor axes, and shows that the effect of GEC heating on planetary radius becomes important for $a > 0.1 \rm~AU$, and is capable of inflating planets to radii consistent with those observed. Note that variations in (a) stellar magnetic field, (b) planetary magnetic field, and c) planetary composition will have potentially significant impacts which are not explored in detail here (though see \cite[Buzasi 2013]{Buzasi13} for limited discussion). Further improvements to the model are planned, and include examining the effects of variations in composition and magnetic field, as well as GEC model interactions with other heating mechanisms, such as Ohmic (\cite[Batygin et al. 2011]{Batygin11}) and tidal heating.

\vspace*{-0.25 cm}
\section{Acknowledgments}

I am grateful to the Whitaker Foundation for its support of the Whitaker Chair at Florida Gulf Coast University. Funding for the Kepler mission is provided by the NASA Science Mission Directorate, and this work was in part funded by the Kepler Participating Science Program grant NNH09CE70C.

\end{document}